\documentclass{article}

\usepackage{graphicx}

\begin{document}

% Title portion
\title{Ion-acoustic eigenmodes in a helical magnetic mirror}

\author{I.S. Chernoshtanov \thanks{Budker Institute of Nuclear Physics, 630090 Novosibirsk,
Russia, I.S.Chernoshtanov@inp.nsk.su}}
%\eaddress{I.S.Chernoshtanov@inp.nsk.su}

%\affiliation{Budker Institute of Nuclear Physics, 630090 Novosibirsk, Russia}

%\affil[aff1]{Budker Institute of Nuclear Physics, 630090 Novosibirsk, Russia}
%\corresp[cor1]{Corresponding author: chernoshtanov@inp.nsk.su}

\maketitle

\begin{abstract}
Plasma rotation together with corrugation of magnetic field can
result in coupling acoustic waves with different azimuthal
wavenumbers and formation of eigenmodes with discrete frequency
spectrum and zero longitudinal group velocity in a helical magnetic
mirror. Such eigenmodes can be destabilized by plasma rotation as
well as resonant interaction with trapped ions. Equations describing
these waves in the linear approximation are derived and results of
numerical solving the equations are discussed. Probably these
eigenmodes may drive anomalous ions scattering which is needed for
effective suppressing flow of rarefied plasma through a helical
mirror.
\end{abstract}

% Head 1
\section{Introduction}

Helical magnetic mirrors is newly proposed method of suppressing
longitudinal losses from magnetic mirrors \cite{Beklemishev13}. If
plasma rotates (because of $E\times B$ drift) in straight magnetic
field with helical corrugation, longitudinal force influences on the
plasma. Such force can decelerate plasma outflow through a helical
mirror and decreases longitudinal losses \cite{Beklemishev16}.
Magnetic field varies periodically along field line, so some of the
ions are trapped between maxima of the magnetic field. Longitudinal
momentum exchange between the trapped and outflow ions are important
for the helical confinement {\cite{Beklemishev16}}. If plasma is too
dense, the Coulomb collisions can provide necessary level of
momentum exchange, but in the case of hot rarefied plasma anomalous
collisionless mechanisms are required.

Experiments at the SMOLa device \cite{SMOLA16} demonstrate effective
suppression of plasma flow through the helical mirror not only if
Coulomb free path $l$ is of the order of helical pitch $h$
\cite{SMOLA19}, but also in the case of rarefied plasma when $l\gg
h$ \cite{SMOLA20}. Excitation of oscillations with several
eigenfrequencies and continuous frequency spectrum is observed;
probably these oscillations provide anomalous momentum exchange.
Properties of these oscillations in different regimes are accurately
measured \cite{SMOLA24}.

Simple estimation of parameters of unstable fluctuations can be made
by assuming that the fluctuations are driven by the resonant Landau
interaction with trapped ions \cite{SMOLA24,Chernoshtanov22}.
Electric field of a wave with spatial structure ${\bf E}\sim
e^{ik_\|z+im\theta-i\omega t}$ can be constant along trajectory of a
trapped ion (in the simplest case the trajectory is a helical line
$r=\mbox{const}$, $\theta=2\pi z/h+\pi$ \cite{Chernoshtanov21}) if
$\omega=\omega_E(m+k_\|h/(2\pi))$, where $\omega_E$ is frequency of
$E\times B$ rotation. In the experiment oscillations with
$\omega\ll\omega_E$ are observed; these waves can be instable if
$k_\|h\approx 2\pi m$. These relationships between frequency and
longitudinal wavenumber of observed oscillation are proved in the
experiment \cite{SMOLA26} but physical nature of the oscillation
remains unclear.
%({\it Link to the article are required here. What status of article which you submit to Journal of Plasma Physics?})

This paper discusses oscillations specific to plasma rotating in a
helical magnetic field. It is well known that spatial periodicity
can results in formation of gaps in frequency spectrum. An example
is Toroidal Alfv{\'e}n eigenmodes which arise in tokamaks because of
corrugation of magnetic field along field lines and coupling waves
with close frequencies and different poloidal wavenumbers
\cite{Heidbrinkb08}. Corrugation of magnetic field provides the
spatial periodicity and plasma rotation produce correlation of the
oscillations along magnetic surface in a helical mirror. It results
in the formation of eigenmodes with discrete frequency spectrum.

Experimental results \cite{SMOLA24,SMOLA26} show that frequencies of
the observed oscillations are much less than ion-cyclotron
frequency, so we use the magnetohydrodynamic approach for simplicity
(of course more detail consideration will require developing kinetic
models). We assume that plasma pressure is much less than pressure
of magnetic field (in the opposite case the helical confinement
doesn't make sense) and consider electrostatic perturbations only.
Most simple is the case of ion-acoustic waves, at the same time this
case is important because longitudinal component of electric field
is non-zero for these oscillations, so the waves can accelerate of
decelerate ions effectively. The article is organized as follows. In
the second section we briefly consider a plasma flow with helical
symmetry, which early was described by Solovi'ev \cite{Soloviev63}.
In the third section equations describing the ion-acoustic waves are
derived and method of numerical solution of the equations is
described in the forth section. Properties of acoustic eigenmodes
found using numerical solution are discussed in the fifth section.

\section{Unperturbed plasma flow}

Let's consider a stationary plasma flow in magnetic field with
helical symmetry (all variables depend on the radial coordinate $r$
and combination $\theta-kz$, where $\theta$ is the azimuthal angle
and $k\equiv2\pi/h$). The following is a modified and simplified (in
particular, influence of plasma on magnetic field is neglected)
summary of the results obtained by Solov'ev \cite{Soloviev63}.
Plasma density $\rho$, velocity $\bf V$ and magnetic field $\bf B$
satisfy the discontinuity equation, the freezing equation and the
equation of motion
\begin{eqnarray}
\nabla(\rho{\bf V})=0,\quad\nabla\times({\bf V}\times{\bf
B})=0,\quad{\bf V}\times(\nabla\times{\bf
V})=-\nabla\left(w+\frac{V^2}{2}\right)+\frac{{\bf
j}_\perp\times{\bf B}}{c\rho},
\end{eqnarray}
where $w=(\gamma/(\gamma-1))p/\rho$ is enthalpy (which is defined so
that $dw=dp/\rho-TdS$; we assume that $dS=0$ later). We will
consider later that plasma current $\bf j$ is small,
$\nabla\times{\bf B}\approx0$.

It is convenient to introduce vector ${\bf h}=\{0,kr,1\}/(1+k^2r^2)$
with the following properties:
\begin{eqnarray*}
\nabla\cdot{\bf h}=0,\quad\nabla\times{\bf
h}=\frac{2k}{1+k^2r^2}{\bf h},\quad{\bf h}\cdot\nabla(\theta-kz)=0.
\end{eqnarray*}
The magnetic field with the helical symmetry can be written in the
following form
\begin{eqnarray*}
{\bf B}=\nabla A\times{\bf h}+b{\bf h},
\end{eqnarray*}
where $A(r,\theta-kz)=krA_\theta+A_z$ is magnetic flux ($A_\theta$
and $A_z$ are cylindrical components of the magnetic vector
potential). The divergence-free condition $\nabla\cdot{\bf B}=0$ is
satisfied automatically, relation $\nabla\times{\bf B}=0$ results in
the condition $b=\mbox{const}$ and the following equation for the
magnetic flux $A$
\begin{eqnarray*}
\nabla^{h}A=\frac{2kb}{1+k^2r^2},\quad\nabla^{h}A\equiv\Delta
A+\frac{2k}{1+k^2r^2}\frac{\partial A}{\partial r}.
\end{eqnarray*}
Three mutually orthogonal vectors $\bf h$, $\nabla A$ and $\nabla
A\times{\bf h}$ form a basis on which it is convenient to decompose
arbitrary vectors.

It is follows from the freezing equation that ${\bf V}\times{\bf
B}=\nabla\Phi$, where the function $\Phi$ is constant along a
magnetic surface $A=\mbox{const}$, i.e.
$\Phi(r,\theta-kz)=\Phi(A(r,\theta-kz))$. The function $\Phi$ is
proportional to the electrostatic potential. One can write plasma
velocity as combination of electric drift and motion along a field
line:
\begin{eqnarray}
%{\bf V}=\frac{{\bf B}\times\nabla\Phi(A)}{B^2}-\left(\left(\frac{1+k^2r^2}{b}-\frac{b}{B^2}\right)\Phi'(A)-\chi(A)\right){\bf
%B}=\nonumber\\=\frac{1+k^2r^2}{b}\Phi'(A){\bf h}\times\nabla A+\chi(A){\bf B}.\label{equ00}
{\bf V}=\left(\frac{{\bf
B}\times\nabla\Phi(A)}{B^2}+\frac{b}{B^2}\Phi'(A){\bf
B}\right)-\left(\frac{1+k^2r^2}{b}\Phi'(A)-\chi(A)\right){\bf
B}=\nonumber\\=\frac{1+k^2r^2}{b}\Phi'(A){\bf h}\times\nabla
A+\chi(A){\bf B}.\label{equ00}
\end{eqnarray}

It is follows from the equation of motion that combination
\begin{equation}
w+\frac{V^2}{2}=\frac{\gamma}{\gamma-1}\frac{p}{\rho}+\frac{1+k^2r^2}{2b^2}|\nabla\Phi|^2+\frac{\chi^2B^2}{2}\equiv
W(A)\label{equ01}
\end{equation}
depends on the magnetic flux $A$ only.

Finally, the discontinuity equation $\nabla\cdot(\rho{\bf V})=0$ can
be written in the following form
\begin{equation}
{\bf
B}\cdot\nabla\left(\left(\chi(A)-\frac{1+k^2r^2}{b}\Phi'(A)\right)\rho\right)=0,\quad\rho=\frac{Q'(A)}{\chi(A)-(1+k^2r^2)\Phi'(A)/b},\label{euu02}
\end{equation}
where $Q(A)$ is mass flow within a magnetic surface
$A=\mbox{const}$.

The expressions (\ref{equ00}), (\ref{equ01}) and (\ref{euu02}) allow
us to find distributions of density, pressure and plasma velocity at
given magnetic flux $A$, mass flow $Q(A)$, electrostatic potential
$\Phi(A)$ and functions $\chi(A)$ and $W(A)$.

\section{Dispersion relation for ion-acoustic oscillations}

We consider the simplest case of electrostatic oscillations, namely
ion-acoustic waves, when perturbed velocity ${\bf v}_1$ is parallel
to the magnetic field, ${\bf v}_1=\xi_1(r,\theta,z){\bf B}$.

The general form of perturbation of plasma density $\rho_1$ and the
function $\xi_1$ can be found from the following considerations.
Small natural oscillations of a system realize irreducible
representations of a symmetry group of this system. Helical symmetry
means that unperturbed variables do not change during the
transformation $\theta\to\theta+k\lambda$ and $z\to z+\lambda$ for
arbitrary real $\lambda$. The corresponding infinitesimal
transformation is $\hat I=k\partial_\theta+\partial_z$. The
eigenfunctions of the transformation (which realize irreducible
representation) satisfies equation $\hat Ig({\bf r})=\kappa g({\bf
r})$, where $\kappa$ is arbitrary real number. Eigenfunctions are
$g({\bf r})=g(A(r,\theta-kz),\theta-kz)e^{i\kappa z}$. So, we can
choose the functions $\chi_1$ and $\rho_1$ in the following form:
\begin{equation}
\rho_1({\bf r},t)=\rho_p(A,\theta-kz)e^{i\kappa(A)z},\quad\xi_1({\bf
r},t)=\xi_p(A,\theta-kz)e^{i\kappa(A)z},\label{xi00}
\end{equation}
where functions $\rho_p(A(r,\theta-kz),\theta-kz)$ and
$\xi_p(A(r,\theta-kz),\theta-kz)$ are periodical functions of
argument $\theta-kz$.

The linearized discontinuity equation $-i\omega\rho_1+\rho{\bf
V}\cdot\nabla(\rho_1/\rho)+\nabla(\rho\cdot{\bf v}_1)=0$ can be
written in the following form
\begin{eqnarray*}
%-i(\omega-\kappa\omega_E)\frac{\rho_1}{\rho_0}+{\bf %B}\cdot\nabla\left(\frac{\rho_1}{\rho_0}\right)+\frac{1}{\rho_0}{\bf B}\cdot\nabla\left(\rho_0\xi_1\right)=0,
-i(\omega-\kappa\frac{\omega_E}{k})\rho_1+{\bf
B}\cdot\nabla\left(\left(\chi-\frac{1+k^2r^2}{b}\frac{\omega_E}{k}\right)\rho_1+\frac{Q'}{\rho}\xi_1\right)=0.
\end{eqnarray*}
It should be noted that the operator ${\bf B}\cdot\nabla$ can be
written as ${\bf
B}\cdot\nabla\xi_1=-r^{-1}(\partial_rA)(\partial_\theta\xi_1)+i\kappa
B_z\xi_1$ for $\xi_1$ given by expression (\ref{xi00}).

The perturbed motion equation is $({\bf V}\cdot\nabla){\bf
v}_1+({\bf v}_1\cdot\nabla){\bf V}=-\nabla w_1+({\bf j}_1\times{\bf
B})/(c\rho_0)-(\rho_1/\rho_0)({\bf j}\times{\bf B})/(c\rho_0)$,
where $w_1$ is perturbation of the enthalpy. The projection of this
equation onto a field line gives
\begin{eqnarray*}
-i\omega\xi_1B^2-({\bf B}\times\nabla\Phi)\cdot\nabla\xi_1+{\bf
B}\cdot\nabla(w_1+{\bf V}\cdot{\bf v}_1)=0.
\end{eqnarray*}
This expression finally can be written as
\begin{eqnarray*}
-i\left(\omega-\kappa\frac{\omega_E}{k}\right)\xi_1B^2+{\bf
B}\cdot\nabla\left(w_1+\left(\chi-\frac{1+k^2r^2}{b}\frac{\omega_E}{k}\right)\xi_1B^2\right)=0.
\end{eqnarray*}

In the case when compression of plasma is polytropic (i.e.
$p/\rho^\gamma=\mbox{const}$) the enthalpy is
$w=(\gamma/(\gamma-2))p/\rho$ and equation for perturbation of the
enthalpy can be written in the following form
\begin{eqnarray*}
-i\left(\omega-\kappa\frac{\omega_E}{k}\right)\frac{w_1}{\rho^{\gamma-2}}+{\bf
B}\cdot\nabla\left(\left(\chi-\frac{1+k^2r^2}{b}\frac{\omega_E}{k}\right)\frac{w_1}{\rho^{\gamma-2}}+\frac{\gamma
p}{\rho^\gamma}\frac{\rho}{B^2}\xi_1B^2\right)=0.
\end{eqnarray*}

Finally, we obtain a system of equations for $\xi_1B^2$ and
$w_1/\rho^{\gamma-2}$:
\begin{eqnarray}
-i\left(\omega-\kappa\frac{\omega_E}{k}\right)\xi_1B^2+{\bf
B}\cdot\nabla\left(\frac{Q'}{\rho}\xi_1B^2+\rho^{\gamma-2}\frac{w_1}{\rho^{\gamma-2}}\right)=0,\nonumber\\
-i\left(\omega-\kappa\frac{\omega_E}{k}\right)\frac{w_1}{\rho^{\gamma-2}}+{\bf
B}\cdot\nabla\left(\frac{Q'}{\rho}\frac{w_1}{\rho^{\gamma-2}}+\frac{c_s^2}{B^2\rho^{\gamma-2}}\xi_1B^2\right)=0.
\label{equ05}
\end{eqnarray}

%Using relations
%\begin{eqnarray*}
%\xi_1B^2=\frac{B^2V_1Q'/\rho-B^2\rho^{\gamma-2}W_1}{B^2Q'^2/\rho^2-c_s^2},\quad\frac{w_1}{\rho^{\gamma-2}}=\frac{B^2W_1Q'/\rho-(c_s^2/\rho^{\gamma-2})V_1}{B^2Q'^2/\rho^2-c_s^2},
%\end{eqnarray*} we move onto new variables $V_1$ and $W_1$, then
%the equations (\ref{equ05}) will take the form
%\begin{eqnarray*}
%-i\left(\omega-\kappa\frac{\omega_E}{k}\right)\left(\frac{BQ'}{\rho}V_1-B\rho^{\gamma-2}W_1\right)+\left(\frac{B^2Q'^2}{\rho^2}-c_s^2\right)\frac{{\bf
%B}\cdot\nabla V_1}{B}=0,\nonumber\\
%-i\left(\omega-\kappa\frac{\omega_E}{k}\right)\left(\frac{BQ'}{\rho}W_1-\frac{c_s^2}{B\rho^{\gamma-2}}V_1\right)+\left(\frac{B^2Q'^2}{\rho^2}-c_s^2\right)\frac{{\bf
%B}\cdot\nabla W_1}{B}=0.
%\end{eqnarray*}
%The coefficient before the term ${\bf B}\cdot\nabla$ a becomes zero
%when $Q'(A)/\rho(A,\theta-kz)=c_s(A,\theta-kz)$. If the longitudinal
%velocity $({\bf B}/B)\cdot{\bf V}$ is small, this corresponds to the
%condition $\omega_E/k\approx c_s$ when the speed of the traffic
%mirrors in the rotating reference frame is close to the ion acoustic
%velocity.
%
%Eliminating $W_1$ from the first equation, we finally obtain
%\begin{eqnarray*}
%\frac{{\bf
%B}}{B}\cdot\nabla\left(\frac{1}{B\rho^{\gamma-2}}\left(\frac{B^2Q'^2}{\rho^2}-c_s^2\right)\frac{{\bf
%B}\cdot\nabla
%V_1}{B}\right)-\frac{\breve\omega^2}{B\rho^{\gamma-2}}V_1-i\breve\omega
%V_1\frac{{\bf
%B}}{B}\cdot\nabla\frac{Q'}{\rho^{\gamma-1}}-2i\breve\omega\frac{Q'}{\rho^{\gamma-1}}\frac{{\bf
%B}}{B}\cdot\nabla V_1=0.
%\end{eqnarray*}

\section{Numerical scheme}

First, we can re-write the equations (\ref{equ05}) in the following
form
\begin{eqnarray*}
-i\left(\omega-\kappa\frac{\omega_E}{k}\right)\xi_1B^2+\frac{\bf
B}{B_0}\cdot\nabla\left(\frac{B_0Q'}{\rho_0}\frac{\rho_0}{\rho}\xi_1B^2+\rho^{\gamma-2}\frac{w_1B_0}{\rho^{\gamma-2}}\right)=0,\nonumber\\
-i\left(\omega-\kappa\frac{\omega_E}{k}\right)\frac{w_1B_0}{(\rho/\rho_0)^{\gamma-2}}+\frac{\bf
B}{B_0}\cdot\nabla\left(\frac{B_0Q'}{\rho_0}\frac{\rho_0}{\rho}\frac{w_1}{(\rho/\rho_0)^{\gamma-2}}+\frac{c_s^2}{(\rho/\rho_0)^{\gamma-2}}\frac{B_0^2}{B^2}\xi_1B^2\right)=0,
\end{eqnarray*}
where $B_0$ and $\rho_0$ are normalizing constants. So
eigenfrequencies $\omega$ depend only on ion sound velocity $c_s$,
normalized plasma mass flow $B_0Q'/\rho_0$, frequency of $E\times B$
rotation, spatial distribution of normalized magnetic field ${\bf
B}/B_0$ and normalized plasma density $\rho/\rho_0$.

We use following method of numerical solving the equations
(\ref{equ05}). Let's introduce a grid along $\theta$;
$\theta_i=-\pi+(i-1)\delta\theta$ for $1\leq i\leq n_t$, where
$\delta\theta=2\pi/(n_t-1)$. Let's introduce coordinates of magnetic
surface $r_i$ such that $A(r_i,\theta_i)=\mbox{const}=A_0$. We use
finite differences and replace the operator ${\bf B}\cdot\nabla$
with the matrix
\begin{eqnarray*}
D_{i,j}=-i\kappa
B_{z,i}\delta_{i,j}+\left(\frac{B_{\theta,i}}{r_i}-kB_{z_i}\right)\left(\frac{\delta_{i,j+1}-\delta_{i,j-1}}{2\delta\theta}
+\frac{\delta_{i,n_t}\delta_{j,1}-\delta_{i,1}\delta_{j,n_t}}{2\delta\theta}\right),
\end{eqnarray*}
where $B_{\theta,i}=B_\theta(r_i,\theta_i)$ and
$B_{z,i}=B_z(r_i,\theta_i)$. Additional terms (which are
proportional to $\delta_{i,1}$ and $\delta_{j,1}$) are needed to
ensure the periodicity condition
$\xi_p|_{\theta=-\pi}=\xi_1|_{\theta=-\pi}$ and
$w_p|_{\theta=-\pi}=w_1|_{\theta=-\pi}$.

Spatial distribution of $\xi_p$ and $w_p$ and shifted
eigenfrequencies $\omega-\kappa\omega_E/\kappa$ can be found as
eigenvectors and eigenfrequencies of the block matrix
\begin{eqnarray*}
\left(\begin{array}{cc} M^{(dia)} & M^{(ru)} \\ M_{i,j}^{(ld)} &
M^{(dia)} \end{array}\right),\quad
M_{i,j}^{(dia)}=D_{i,j}\frac{Q'}{\rho_j},\quad
M_{i,j}^{(ru)}=D_{i,j}\rho_j^{\gamma-2},\quad
M_{i,j}^{(ld)}=D_{i,j}\frac{c_{s,j}^2}{B_j^2\rho_j^{\gamma-2}}.
\end{eqnarray*}

In this scheme, the values of the function at non-neighboring points
are used to calculate the derivative. It results in arising
non-physical eigenmodes with small-scale eigenvectors. To reject
these modes, all modes for which the number of maxima exceeded 10
are ignored.

\section{Numerical results}

We choose magnetic flux $A(r,\theta,z)$ in the simplest form
\begin{eqnarray*}
A(r,\theta,z)=\frac{kbr^2}{2}+\beta brI_1'(kr)\cos(\theta-kz),
\end{eqnarray*}
where $I_1(x)$ is modified Bessel function of the first kind and the
prime denotes the derivative with respect to the argument. The first
term corresponds to the uniform magnetic field $b{\bf e}_z$ and the
second one is addition of the helical field. The unperturbed
electrostatic potential is chosen in the simplest form
$\Phi(A)=(\omega_E/k)A$, where $\omega_E=\mbox{const}$ is the solid
state plasma rotation frequency.

Let's introduce $r_\star$, which is radial coordinate of magnetic
surface $A(r,\theta-kz)=\mbox{const}$ at $r=r_\star$ and
$\theta=kz$. Unperturbed distributions of plasma density $\rho$ and
local ion acoustic velocity $c_s=\sqrt{\gamma p/\rho}$ at given
magnetic surface are
\begin{eqnarray*}
\rho=\rho_\star\frac{b\chi(A)-(1+k^2r_\star^2)\omega_E/k}{b\chi(A)-(1+k^2r^2)\omega_E/k},\quad
c_s^2=c_\star^2+\frac{\gamma-1}{2}\left((r_\star^2-r^2)\omega_E^2+(B_\star^2-B^2)\chi^2\right),
\end{eqnarray*}
where $B_\star=B(r_\star,kz,z)$,
$\rho_\star\equiv\rho(r_\star,kz,z)$ and $c_\star=\sqrt{\gamma
p(r_\star,kz,z)/\rho_\star}$. We choose following basic parameters:
the helical pitch $h=18$ cm (so that $k=2\pi/h=\pi/9$ cm$^{-1}$) and
radius of magnetic surface $r_\star=3$ cm. Hereinafter we normalizes
all velocities on $c_\star$ and magnetic field on $b$, so that
$c_\star=1$ and $b=1$.

\begin{figure}
\includegraphics[width=0.32\textwidth]{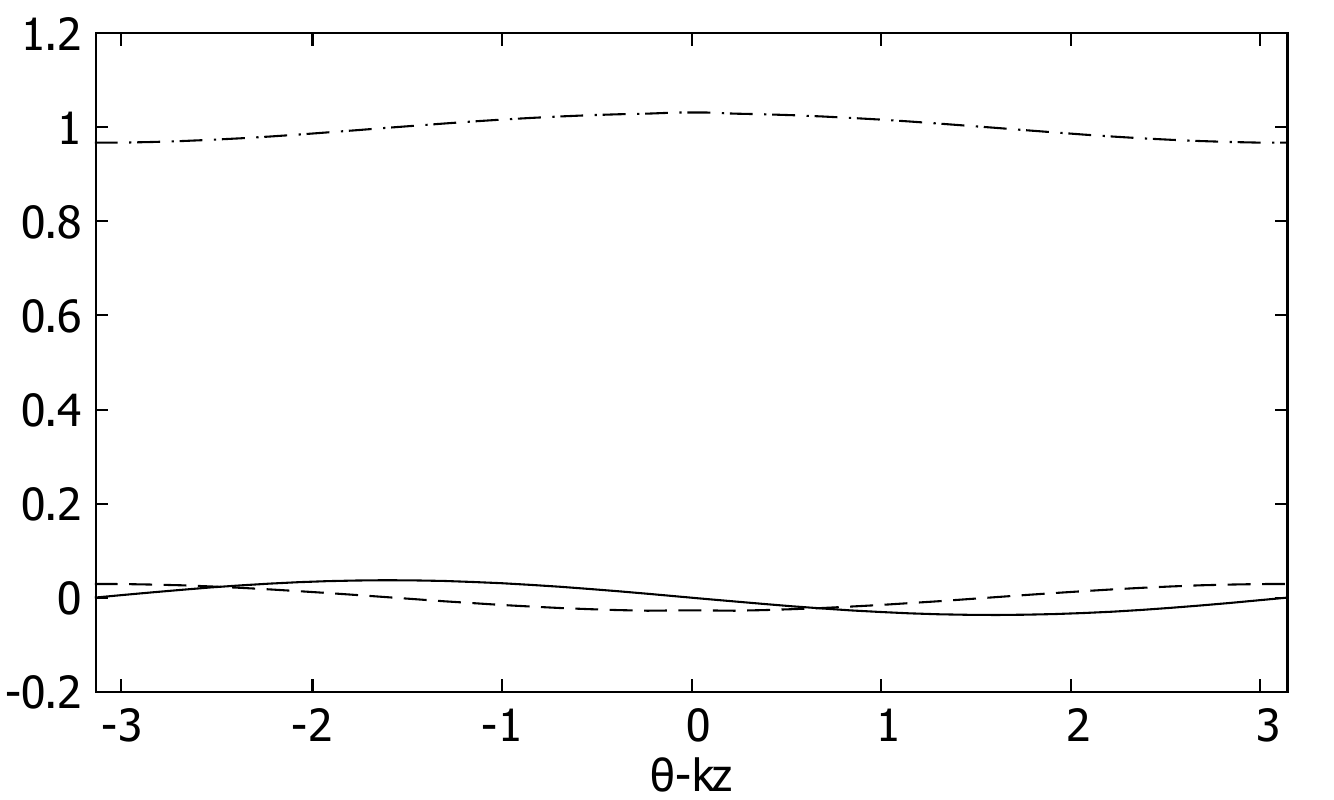}
\includegraphics[width=0.32\textwidth]{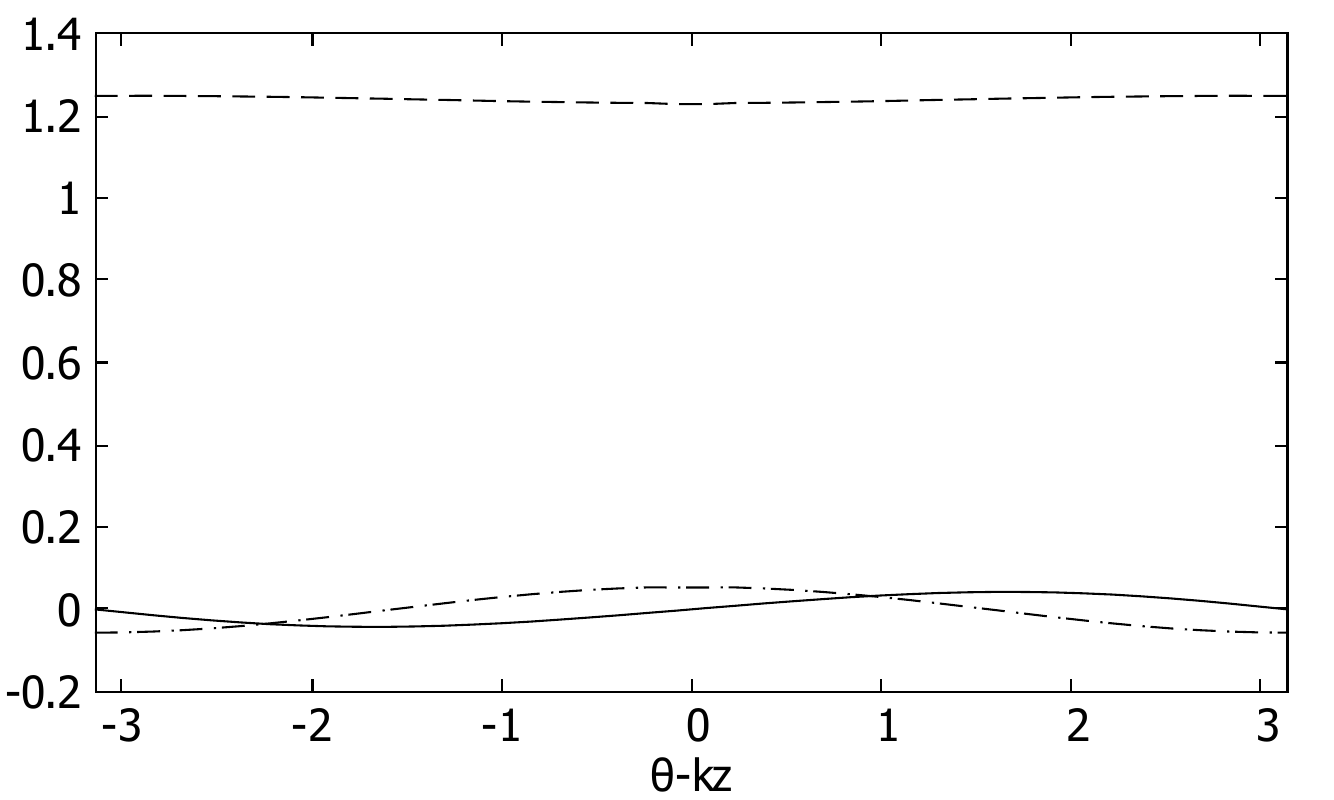}
\includegraphics[width=0.32\textwidth]{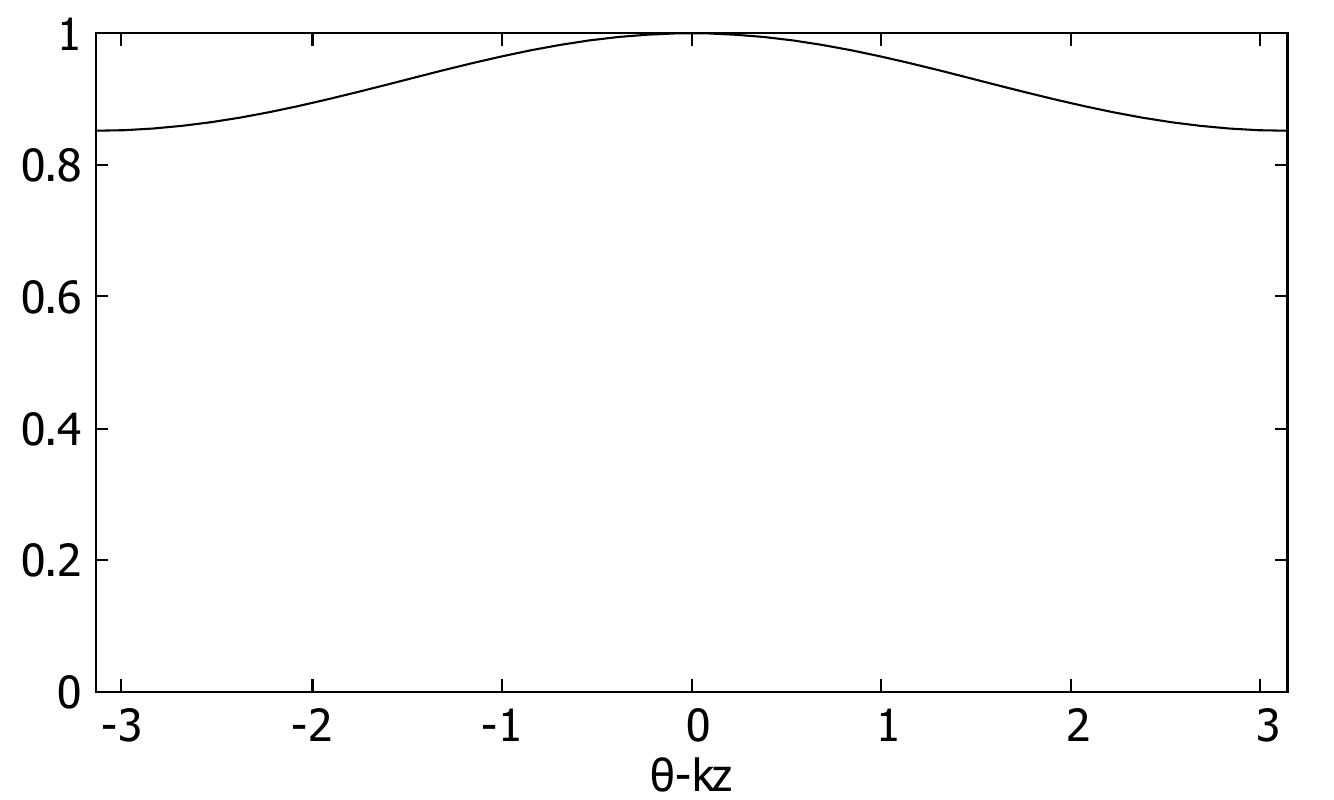}
\caption{Radial (solid line), azimuthal (dashed line) and
$z$-component of the magnetic field (left), components of normalized
unperturbed plasma velocity ${\bf V}/c_\star$ (medium) and
normalized unperturbed plasma density (right). Parameters:
$\beta=0.05$, $r_\star=3$ cm, $\omega_Er_\star/c_\star=1.2$, other
parameters given in the text.} \label{BvR}
\end{figure}

An example of spatial distributions of components of magnetic field,
unperturbed plasma density and velocities are shown in figure
\ref{BvR}. The variable $\chi(A)=2\omega_EA/b^2$ is chosen so that
unperturbed longitudinal plasma velocity $V_\|\equiv({\bf
B}/B)\cdot{\bf V}$ approximately equal to zero. Plasma density is
peaked at surface $\theta=kz$, where magnitude of the magnetic field
is maximal. At $\theta=kz$ magnetic surfaces $A=\mbox{const}$ are
moving closer together and plasma density rises due to conservation
of plasma mass flow.

\begin{figure}
\includegraphics[width=0.5\textwidth]{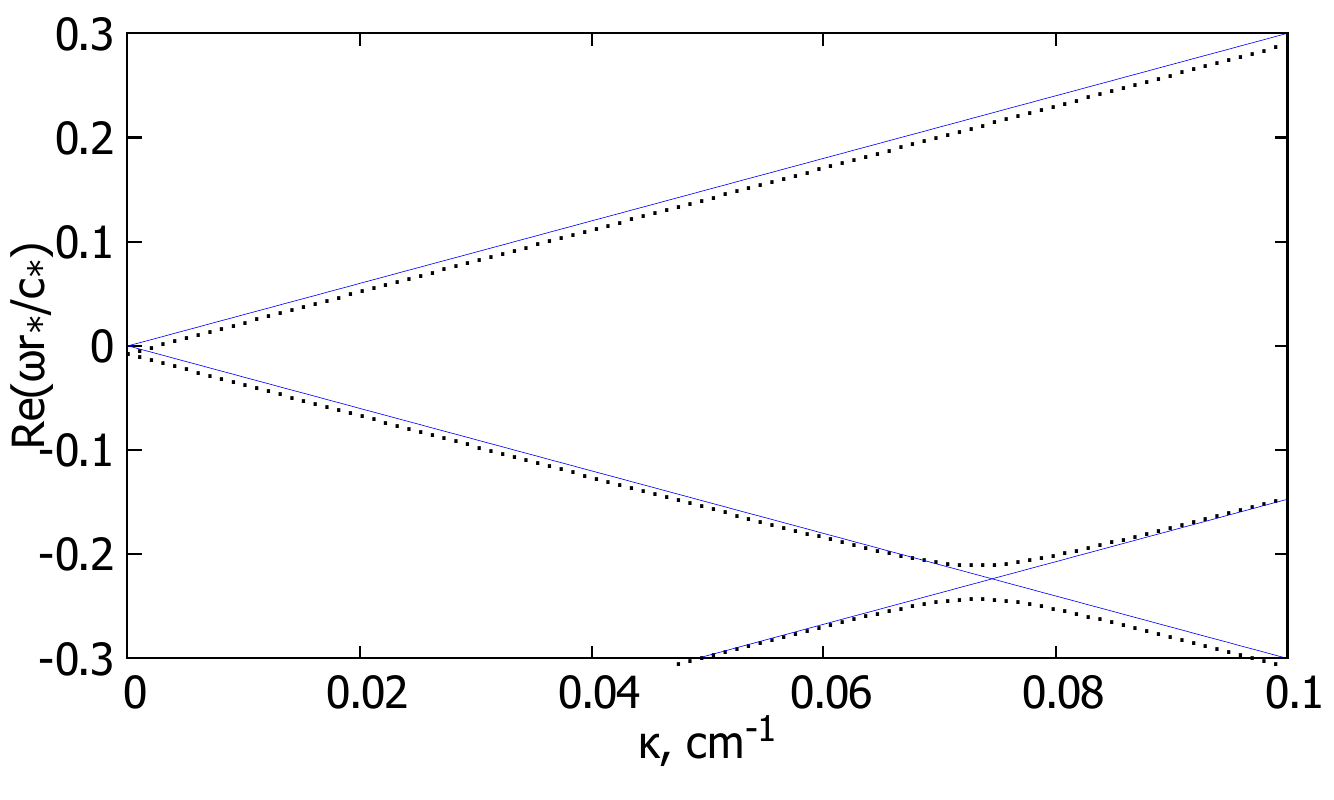}
\includegraphics[width=0.5\textwidth]{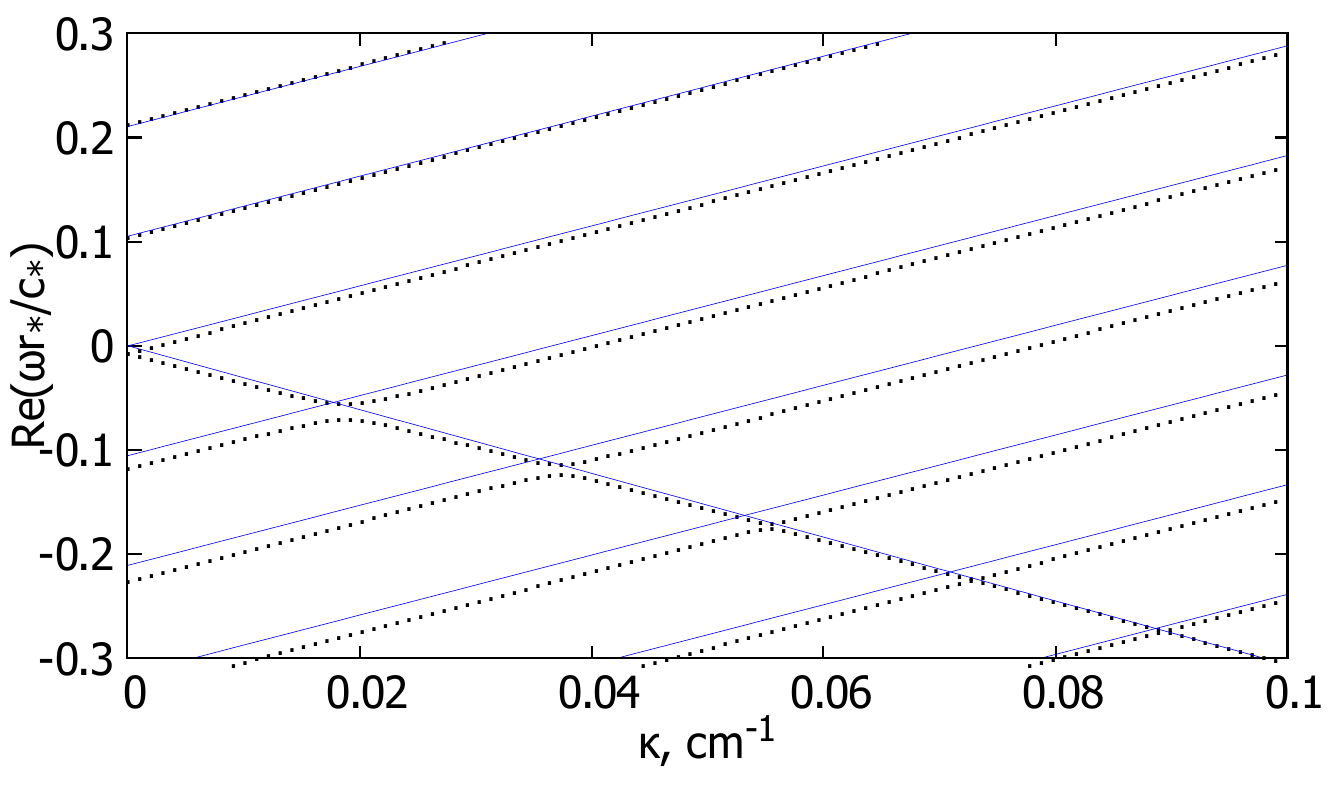}
\includegraphics[width=0.5\textwidth]{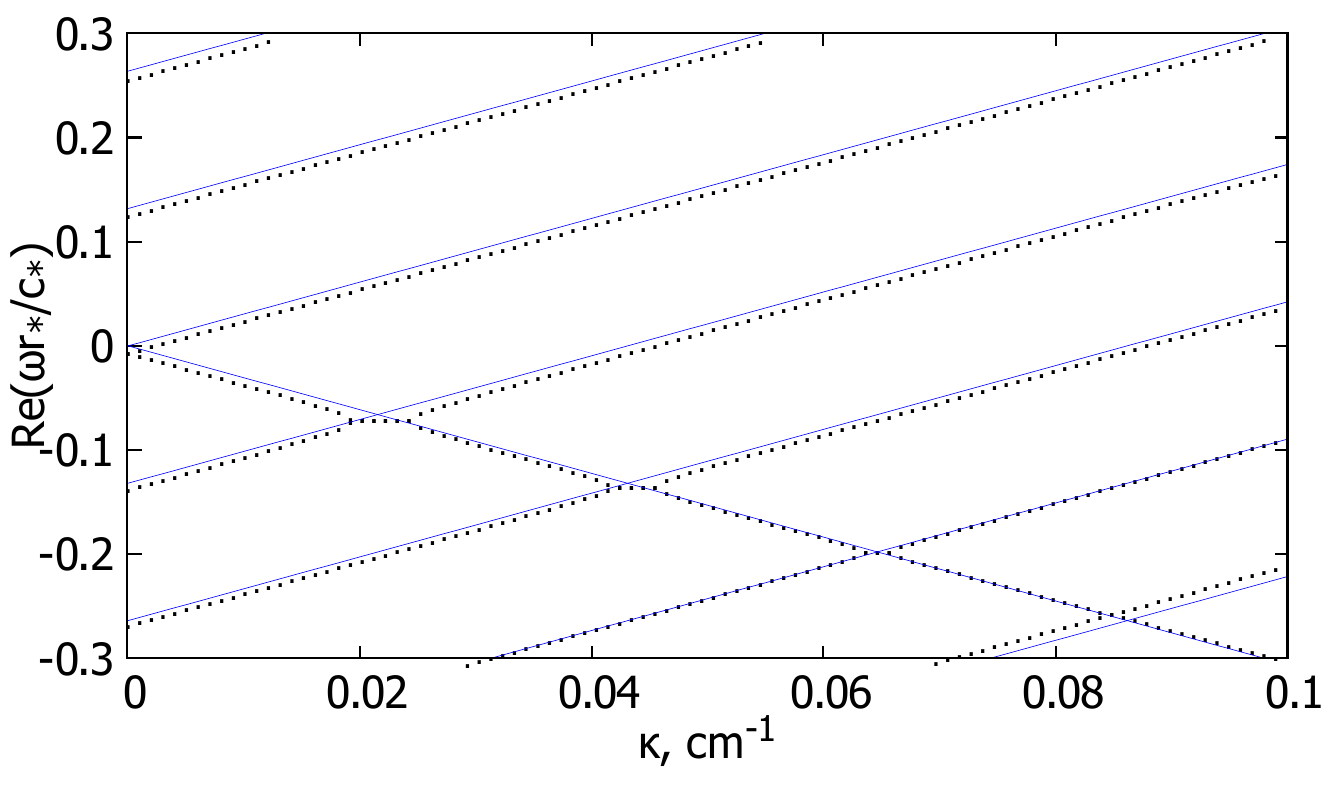}
\includegraphics[width=0.5\textwidth]{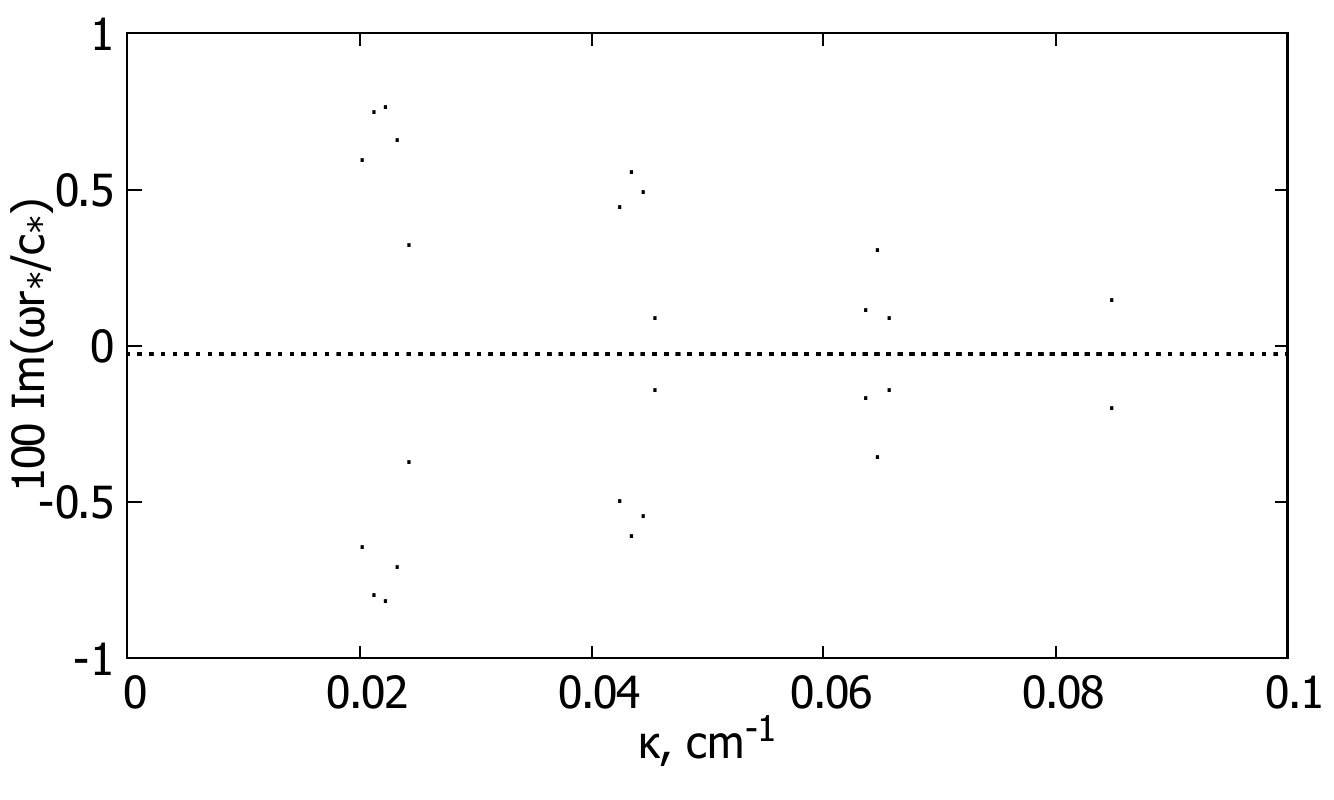}
\caption{Evolution of dispersion curves (points) during increasing
frequency of electric drift. Real part of the frequency at
$\omega_Er_\star/c_\star=0.6$ (left up),
$\omega_Er_\star/c_\star=0.9$ (right up),
$\omega_Er_\star/c_\star=1.2$ (left down) and imaginary part of the
frequency at $\omega_Er_\star/c_\star=1.2$ (right down). Solid lines
are approximate expressions for uniform plasma
$\omega=m\bar\omega_E+(\kappa-mk)(\bar V_z\pm\bar c_s)$. Approximate
values of averaged variables are $\hat\omega_E=\omega_E$, $\hat
V_\|\approx0$, $\bar c_s=c_\star$ in the first case, $\hat
c_s=0.96c_\star$ in the second case and $\hat c_s=1.02c_\star$ in
the second case. Parameters are the same as in figure \ref{Omega}.}
\label{Omega}
\end{figure}

An example of dependence of frequencies on the ``longitudinal
wavevector'' $\kappa$ is shown in figure \ref{Omega}. Magnetic field
corrugation is relatively weak ($B_r/B_z\sim B_\theta/B_z\sim0.05$)
and spatial distribution of oscillations is $\xi_1\approx e^{i\kappa
z}e^{im(\theta-kz)}$ (except oscillations with some frequencies, see
below), where $m$ is azimuthal wavenumber. The dispersion curves are
described approximately by expression
$\omega-m\bar\omega_E=(\kappa-mk)(\bar V_\|\pm\bar c_s)$, where
$\bar\omega_E$, $\bar c_s$ and $\bar V_\|$ are effective frequency
of electric drift, ion-acoustic speed and mean longitudinal velocity
at given magnetic surface. If corrugation is small, these effective
parameters are $\bar\omega_E\equiv\left(\int_{-\pi}^\pi
rd\theta/(2\pi V_\theta)\right)^{-1}$, $\bar
c_s\equiv\bar\omega_E\int_{-\pi}^\pi c_srd\theta/(2\pi V_\theta)$
and $\bar V_\|=\bar\omega_E\int_{-\pi}^\pi V_zrd\theta/(2\pi
V_\theta)$; the analytical expressions can easily be derived from
the dispersion equation for acoustic waves in a
longitudinally-uniform plasma
$\omega-V_\theta(\theta)k_\theta(\theta)=(\kappa-mk)(V_z(\theta)\pm
c_s(\theta))$ and the condition of periodicity $\int_{-\pi}^\pi
k_\theta rd\theta=2\pi m$. Rotation induced by electric drift
removes plasma between different field lines and (together with
corrugation of magnetic field) provokes coupling modes with
different wavenumbers $m$. It results in formation of modes with
non-trivial spatial structure and zero longitudinal group velocity
in regions where the dispersion curves intersect. Frequency of such
helical acoustic modes can be found from condition of coincidence of
frequencies of waves with azimuthal numbers $n$ and $m$:
$\omega_{n,m}=m\bar\omega_E-(\kappa_{n,m}-mk)(\bar V_\|+\bar
c_s)=n\bar\omega_E-(\kappa_{n,m}-nk)(\bar V_\|-\bar c_s)$. It
follows from this condition that
\begin{eqnarray*}
\omega_{n,m}=\frac{n+m}{2}\bar\omega_E+\frac{n-m}{2}\frac{\bar
V_\|}{\bar c_s}\bar\omega_E-\frac{n-m}{2}\left(1-\frac{\bar
V_\|^2}{\bar c_s^2}\right)k\bar
c_s,\nonumber\\
\kappa_{n,m}=\frac{(m-n)\bar\omega_E+(m+n)k\bar c_s+(m-n)k\bar
V_\|}{2\bar c_s}
\end{eqnarray*}
When plasma rotation is fast enough (velocity $B_0Q'/\rho_0$ exceeds
local ion sound speed),
%the dispersion curves coincide and
these eigenmodes become instable. Probably the modes can be
destabilized by trapped ions even if plasma flow is subsonic.

\begin{figure}
\includegraphics[width=0.48\textwidth]{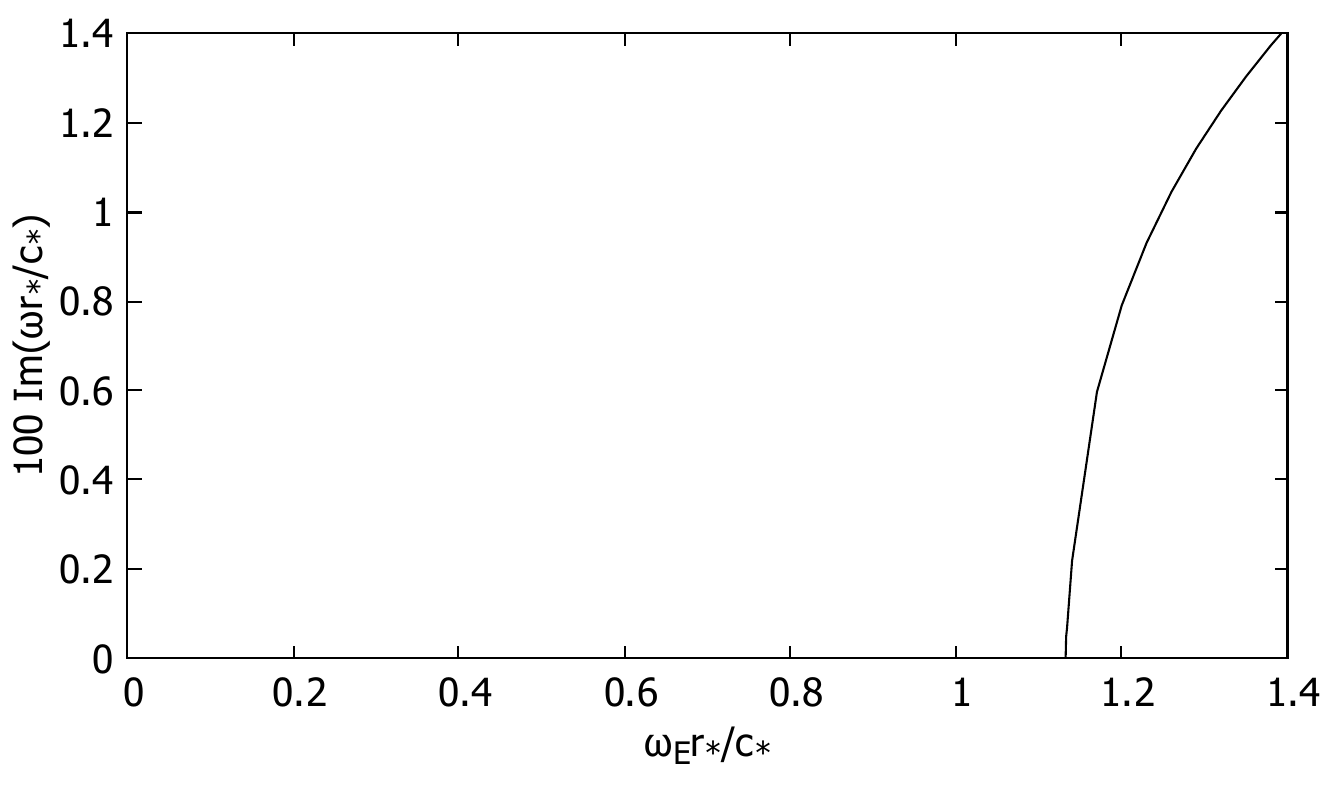}
\includegraphics[width=0.48\textwidth]{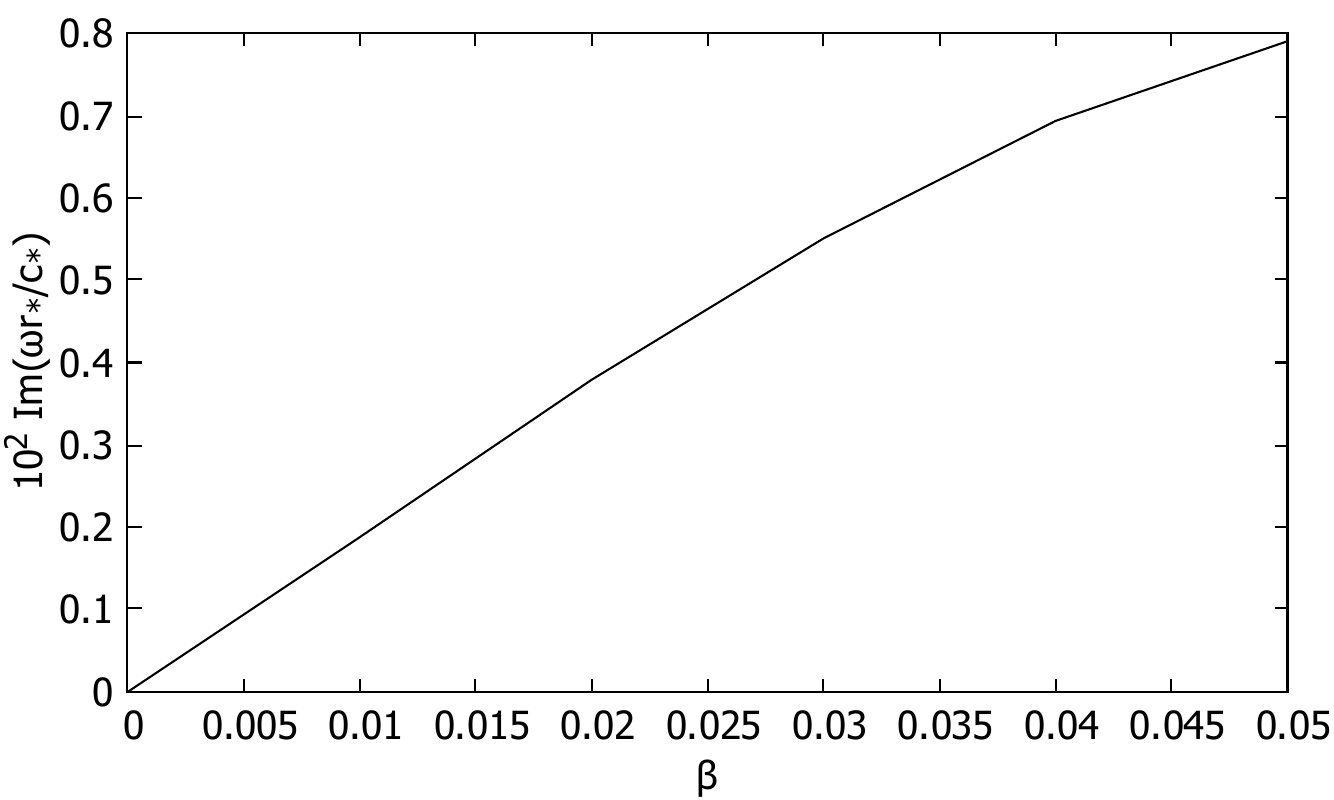}
\caption{Dependence of maximal increment (for $0<\kappa<0.05$
cm$^{-1}$) on frequency of electric drift (left) and helical
corrugation (right). Parameters are the same as in figure
\ref{BvR}.} \label{GammaOB}
\end{figure}

Increment of instability depends on frequency of rotation $\omega_E$
as $\mbox{const}(\omega_E-\omega_{Ec})^{1/2}$, where $\omega_{Ec}$
is critical frequency of rotation (see figure \ref{GammaOB}, left).
Such dependence is typical in the case when instability threshold is
exceeded insufficiently.
% (so-called soft excitation of instability)
It is interesting that there are no threshold when corrugation of
the magnetic field varies. The instability increment simply rises
monotonically during corrugation increasing (see figure
\ref{GammaOB}, right).

\begin{figure}
\includegraphics[width=0.5\textwidth]{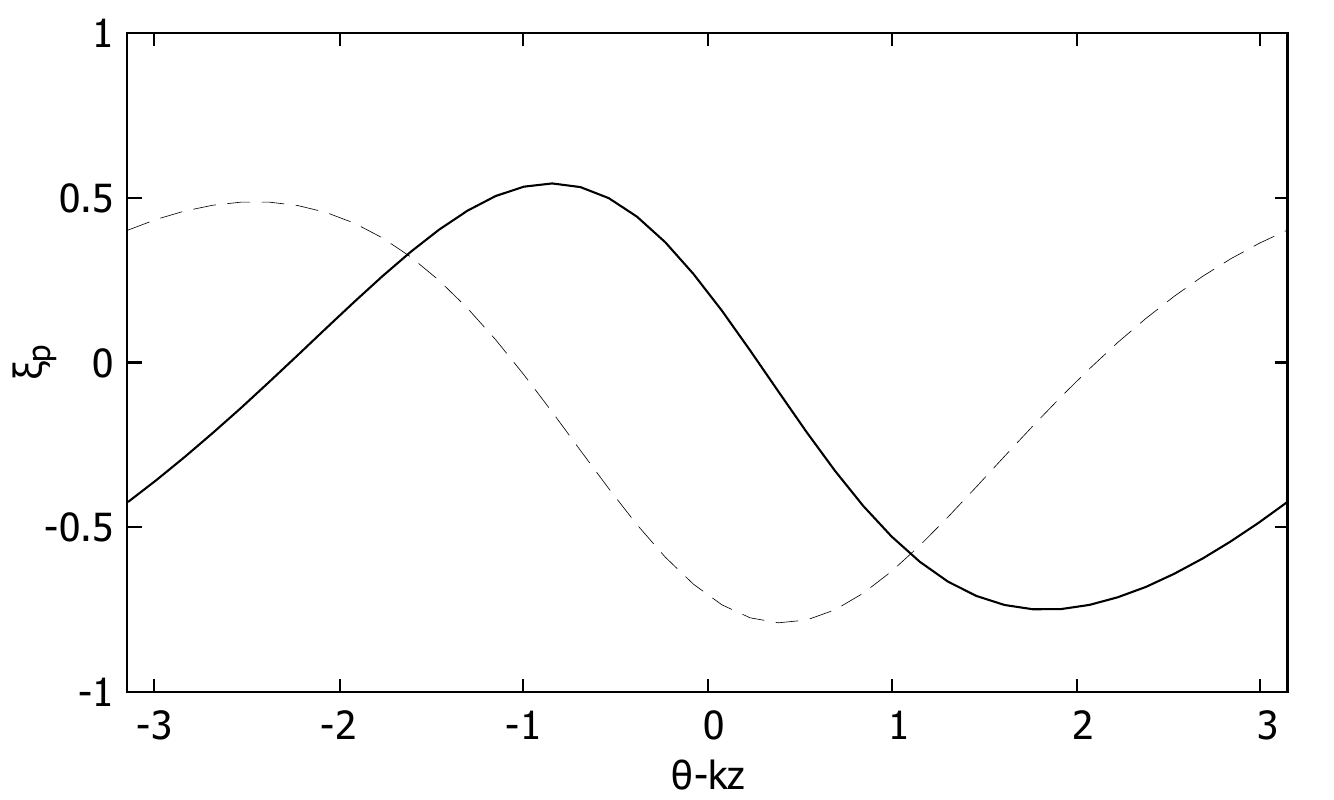}
\includegraphics[width=0.5\textwidth]{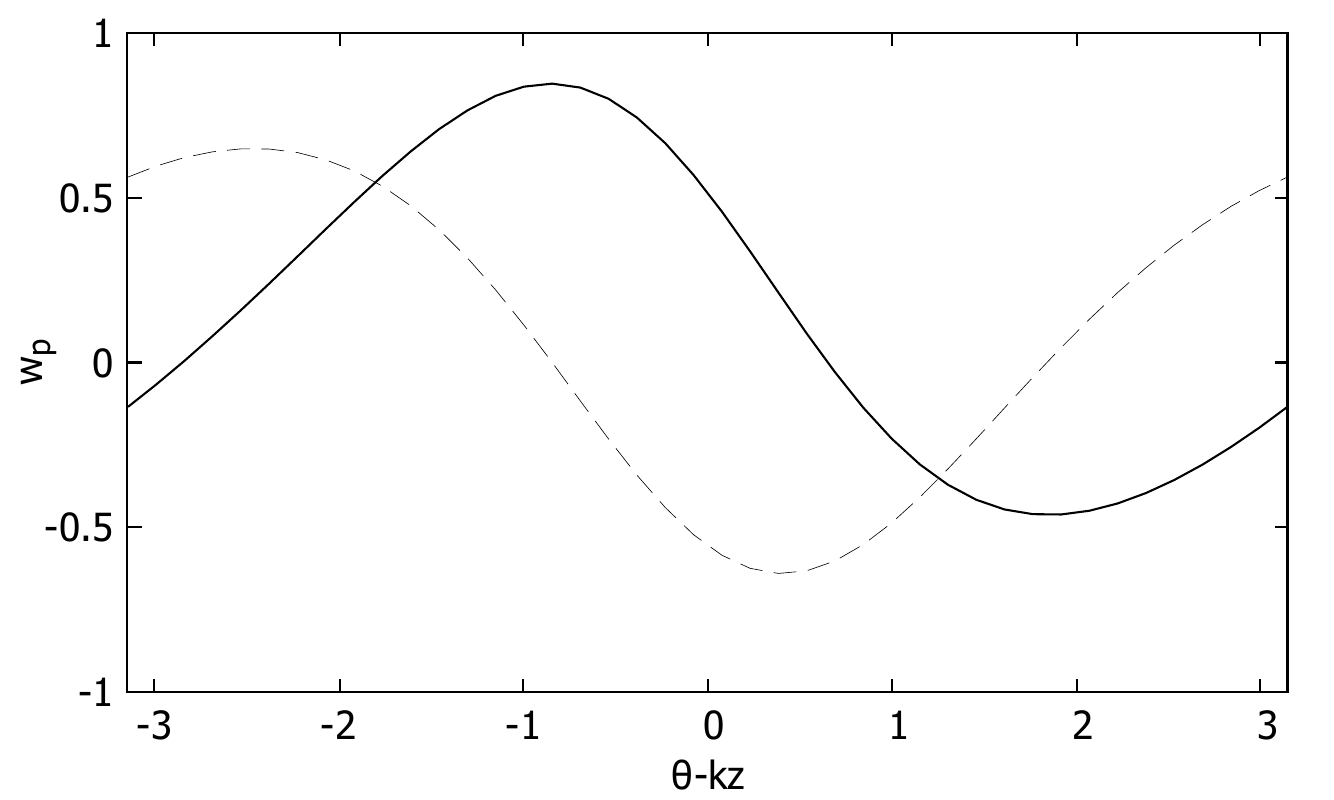}
\includegraphics[width=0.5\textwidth]{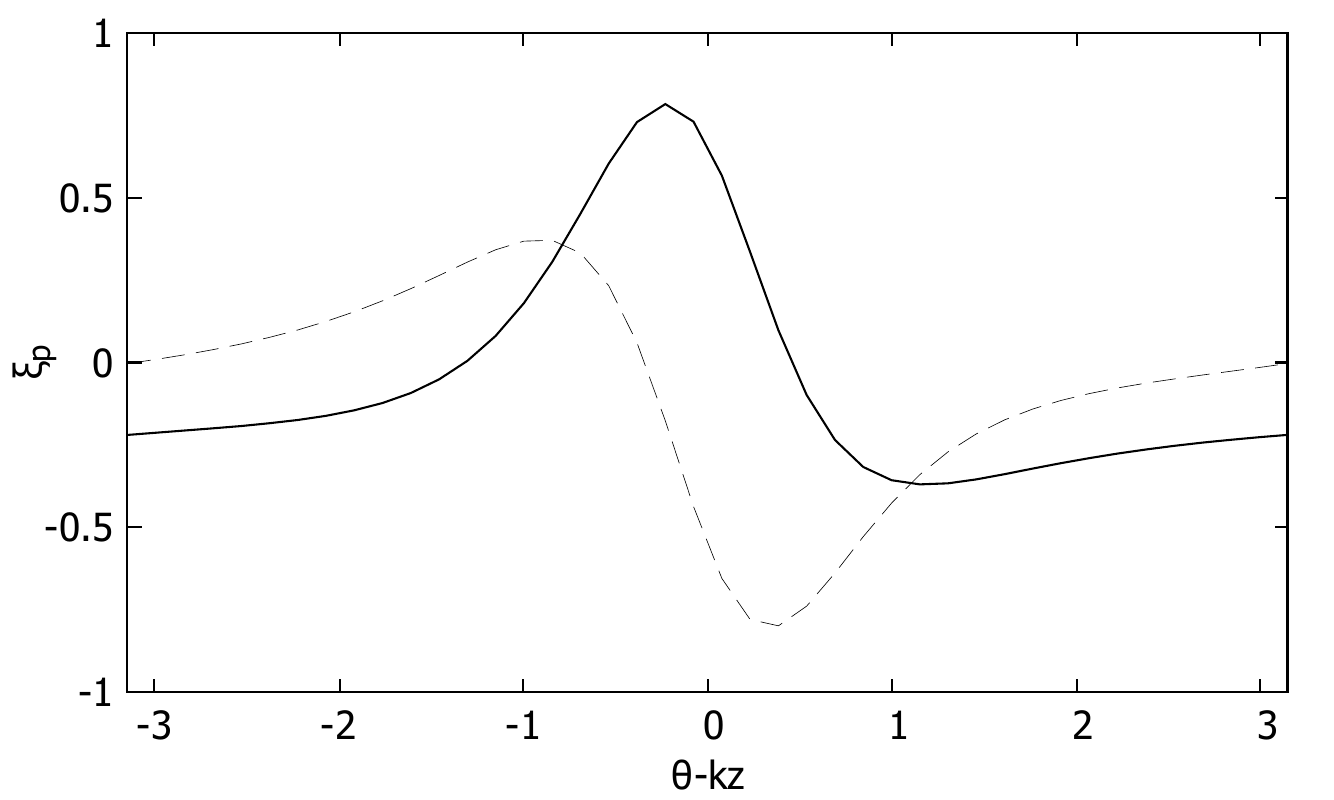}
\includegraphics[width=0.5\textwidth]{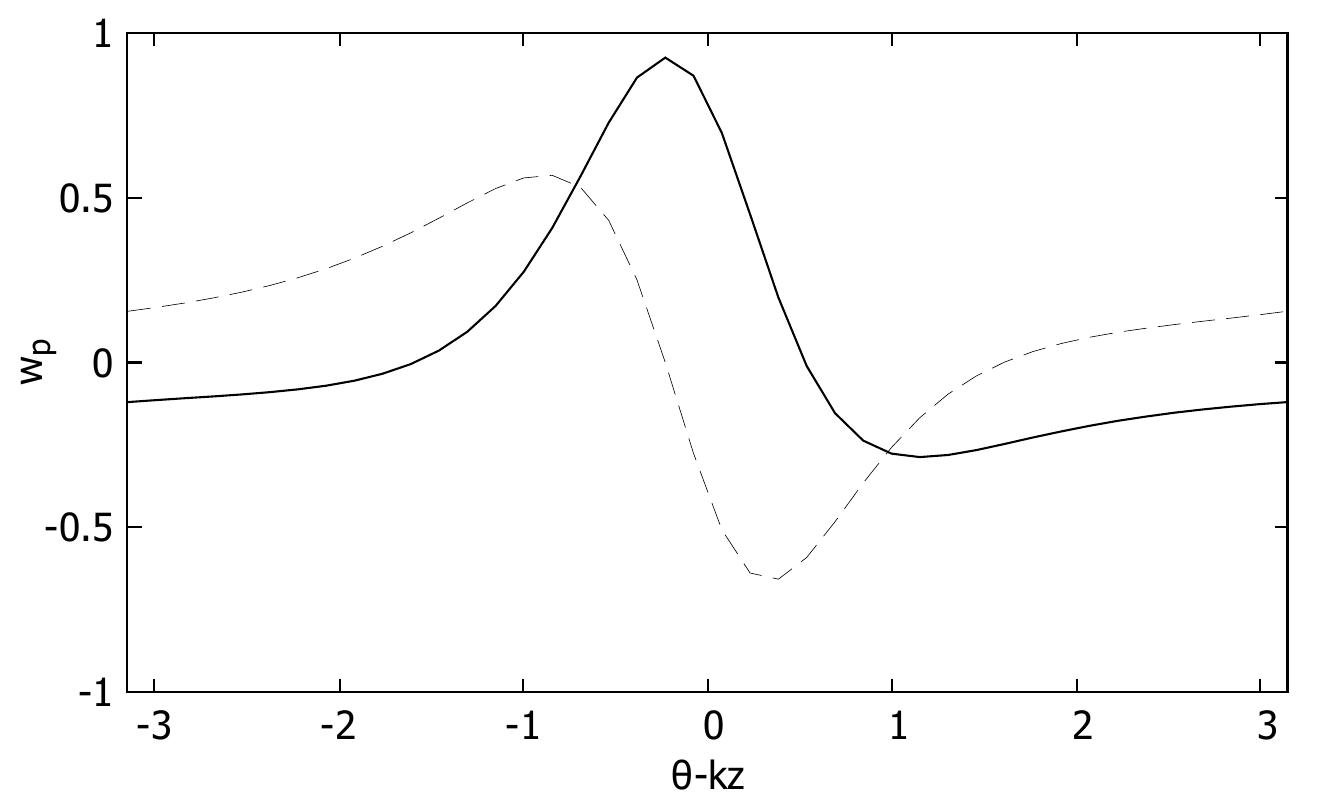}
\caption{Dependence of real (solid lines) and imaginary (dashed
lines) parts of periodical part of perturbation of velocity $\xi_p$
(left) and perturbation of enthalpy $w_p$ (right) on $\theta-kz$.
Helical corrugation is $\beta=0.01$ (up) and $\beta=0.05$ (down),
$\kappa=0.022$, other parameters are the same as in figure
\ref{BvR}.} \label{XiWp01}
\end{figure}

\begin{figure}
\includegraphics[width=0.48\textwidth]{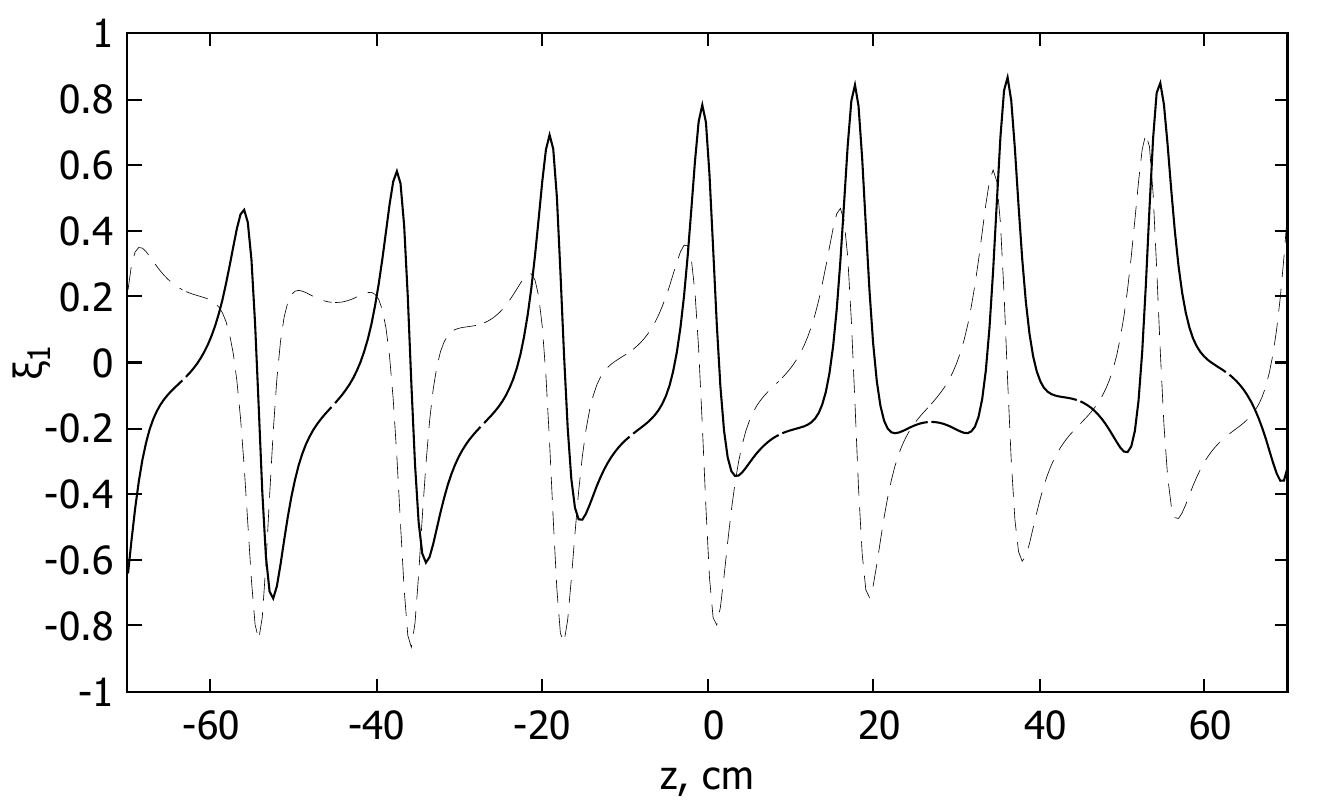}
\includegraphics[width=0.48\textwidth]{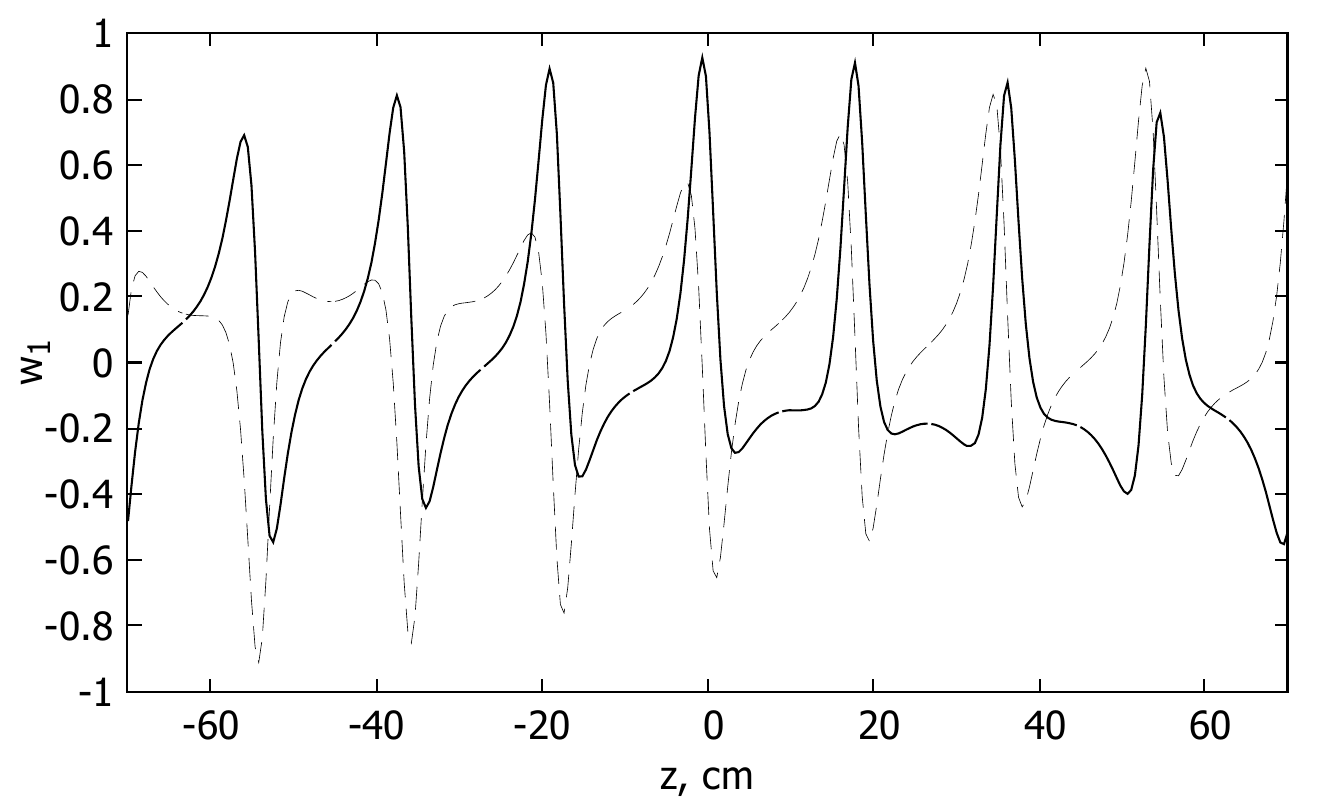}
\caption{Dependence of perturbation of velocity $\xi_1$ (left) and
perturbation of enthalpy $w_1$ (right) on longitudinal coordinate
$z$. Parameters are the same as in down figure \ref{XiWp01}.}
\label{XiW01}
\end{figure}

Spatial distribution of perturbed velocity and enthalpy is
non-trivial for these eigenmodes. Periodical parts
$\xi_p(A,\theta-kz)$ and $w_p(A,\theta-kz)$ are sums of two
azimuthal modes (see figure \ref{XiWp01} where mode numbers are
$m=0$ and $n=1$). Increasing helical corrugation results in
sophisticating spatial structure of the eigenmodes. Because of the
factor $e^{i\kappa z}$ dependence of the perturbed variables $\xi_1$
and $w_1$ on the longitudinal coordinate $z$ is non-periodical (see
figure \ref{XiW01}).

\section{Conclusion}

Plasma rotation in combination with helical corrugation of magnetic
field can result in formation of acoustic waves with non-trivial
spatial structure and zero longitudinal group velocity
$\partial\omega/\partial\kappa$ in a helical magnetic mirror. These
modes arise because of coupling usual acoustic waves with different
azimuthal wavenumbers in the same manner as Toroidal Alfv{\'e}n
eigenmodes arise a tokamak because of coupling the Alfv{\'e}n waves
with different poloidal wavenumbers. Frequencies of such helical
acoustic modes satisfy condition of intersection of dispersion
curves of acoustic waves with different azimuthal wavenumbers $n$
and $m$; this condition gives set of eigenfrequencies
$\omega_{n,m}(A)$ for each magnetic surface $A=\mbox{const}$.
Spatial structure of these modes is non-trivial, in particular there
are no define azimuthal wavenumber and longitudinal periodicity. The
oscillation at different magnetic surfaces are independent in the
MHD approximation; it may results in ``continuum damping'' at all
magnetic surfaces except surfaces where
$\partial\omega_{m,n}/\partial A=0$. Radial structure of these modes
requires additional consideration.

The acoustic eigenmodes can be destabilized by plasma rotation (and
longitudinal plasma outflow) as well as by the resonant Landau
interaction with trapped ions. Plasma rotation probably should be
supersonic for destabilization. From the other hand, because of zero
group velocity such modes may easily be excited by trapped ions even
if plasma flow is subsonic. Detail consideration of interaction of
trapped ions with such modes will probably require developing
kinetic models. Because phase velocity of these modes is similar to
the ion acoustic velocity and thermal velocity of ions, the modes
may effectively change longitudinal velocity of ions and influence
on plasma dynamic in a helical magnetic mirror.

% Acknowledgement
\section{ACKNOWLEDGMENTS}

Author is pleased to Dr. A.D. Beklemishev, Dr. A.V. Sudnukov, Mr.
M.S. Tolkachev and Dr. D.I. Skovorodin for fruitful discussions.


\begin{thebibliography}{99}

\bibitem{Beklemishev13}
%\refitem{article}
{\it A.D. Beklemishev.} Helicoidal System for Axial Plasma Pumping
in Linear Traps // Fusion Science and Technology, 63:1T, 355-357,
(2013), doi: 10.13182/FST13-A16953

\bibitem{Beklemishev16}
%\refitem{article}
{\it A.D. Beklemishev.} Radial and Axial Transport in Trap Sections
with Helical Corrugation // AIP Conference Proceedings {\bf 1771},
040006 (2016), doi: 10.1063/1.496419

\bibitem{SMOLA16}
{\it A.V. Sudnikov, A.D. Beklemishev, V.V. Postupaev, A.V. Burdakov,
I.A. Ivanov, N.G. Vasilyeva, K.N. Kuklin, A.G. Makarov, and E.N.
Sidorov.} Helical mirror concept exploration: Design and status
// AIP Conference Proceedings {\bf 1771}, 030002 (2016); doi:
10.1063/1.4964158

\bibitem{SMOLA19}
{\it A.V. Sudnikov, A.D. Beklemishev, V.V. Postupaev, I.A. Ivanov,
A.A. Inzgevatkina, V.F. Sklyarov, A.V. Burdakov, K.N. Kuklin, A.F.
Rovenskikh and N.A. Melnikov.} First Experimental Campaign on SMOLA
Helical Mirror // Plasma and Fusion Research: Regular Articles 14,
2402023 (2019) doi: 10.1585/pfr.14.2402023

\bibitem{SMOLA20}
{\it A.V. Sudnikov, A.D. Beklemishev, A.A. Inzhevatkina, I.A.
Ivanov, V.V. Postupaev, A.V. Burdakov, V.V. Glinskiy, K.N. Kuklin,
A.F. Rovenskikh and V.O. Ustyuzhanin} Preliminary experimental
scaling of the helical mirror confinement effectiveness // Journal
of Plasma Physics, vol. 86, 905860515 (2020) doi:
10.1017/S0022377820001245

\bibitem{SMOLA24}
{\it M.S. Tolkachev, A.A. Inzhevatkina, A.V. Sudnikov and I.S.
Chernoshtanov.} Electromagnetic oscillations and anomalous ion
scattering in the helically symmetric multiple-mirror trap //
Journal of Plasma Physics, 90(1):975900102 (2024) doi:
10.1017/S0022377823001496

\bibitem{Chernoshtanov22}
{\it I.S. Chernoshtanov.} Influence of low-frequency oscillations on
ions dynamic in an helical mirror. // In proc. of XLIX Zvenigorod
International Conference on Plasma Physics and Controlled Fusion.
Zvenigorod, Russia, March 2022 . doi: 10.34854/ICPAF.2022.49.1.033

\bibitem{Chernoshtanov21}
{\it I.S. Chernoshtanov and D.A. Ayupov.} Collisionless particle
dynamics in trap sections with helical corrugation // Physics of
Plasmas 28, 032502 (2021). doi: 10.1063/5.0040715

\bibitem{SMOLA26}
{\it M.S. Tolkachev, A.V. Sudnikov, A.A. Inzhevatkina, A.V.
Kozhevnikov, V.O. Ustyuzhanin and I.S. Chernoshtanov.} Energy
Spectrum Dynamics in SMOLA Helical Mirror // Journal of Plasma
Physics, submitted for publication

\bibitem{Heidbrinkb08}
%\refitem{article}
{\it W.W. Heidbrinkb.} Basic physics of Alfv{\'e}n instabilities
driven by energetic particles in toroidally confined plasma //
Physics of Plasmas {\bf 15}, pp. 055501, (2008), doi:
10.1063/1.2838239

\bibitem{Soloviev63}
{\it L.S. Solov'ev.} Symmetric magnetohydrodynamic flows and helical
waves in a round plasma cylinder, in {\it Reviews of Plasma
Physics}, Vol. 3,  Authorized translation from the Russian by
Herbert Lashinsky, University of Maryland, USA. Edited by M. A.
Leontovich. Published by Consultants Bureau, New York, 1967, p.277.

\end{thebibliography}
\end{document}